# Theoretical Analysis on Deflagration-to-Detonation Transition[*]

Yun-Feng Liu(刘云峰)[1,2][**], Huan Shen(沈欢)[1,2], De-Liang Zhang(张德良)[1,2], Zong-Lin Jiang(姜宗林)[1,2]

[1] *Institute of Mechanics, Chinese Academy of Sciences, Beijing, 100190*

[2] *School of Engineering Science, University of Chinese Academy of Sciences, Beijing, 100049*

The study on deflagration-to-detonation transition (DDT) is very important because this mechanism has relevance to safety issues in industries, where combustible premixed gases are in general use. However, the quantitative prediction of DDT is one of the major unsolved problems in combustion and detonation theory to date. In this paper, the DDT process is studied theoretically and the critical condition is given by a concise theoretical expression. The results show that a deflagration wave propagating with about 60% Chapman-Jouguet (C-J) detonation velocity is a critical condition. This velocity is the maximum propagating velocity of a deflagration wave and almost equal to the sound speed of combustion products. When this critical conation is reached, a C-J detonation is triggered immediately. This is the quantitative criteria of the DDT process.

## 1. Introduction

It is generally known that combustion waves propagating in a premixed detonable mixture can be classified into two modes, either deflagration or detonation. The detonation wave is initiated by two ways, one is direct initiation, and the other is deflagration-to-detonation transition (DDT). The direct detonation initiation needs extremely high ignition energy supplied to the premixed detonable mixture. Once the detonation is initiated directly, it will propagate self-sustainable with a Chapman-Jouguet (C-J) detonation velocity.

Most of the detonation initiation is imitated by the DDT process. The flame is ignited by a small amount of energy at the initial stage and then the flame front is accelerated by natural or artificial reasons from laminar flame with velocities at the level of some meters per second to turbulent flame with speeds at the level of a hundred meters per second. Under certain critical conditions, the DDT process can occur and the C-J detonation is triggered immediately. There is a critical threshold inherent in the DDT process. Sufficient evidence from the previous studies indicates that the deflagration velocity has to reach some quasi-steady critical value which is relatively repeatable before the DDT process abrupt.

[*] Supported by the National Natural Science Foundation of China (Grants 11672312 and 11532014).

[**] Corresponding author:liuyunfeng@imech.ac.cn

The DDT process has attracted considerable research interest. Over the last half century, many experimental and numerical studies were conducted for detonation onset and DDT process to test a wide range of initial and boundary conditions. The DDT was first observed in experiments by Brinkley and Lewis [1]. Then, Oppenheim and his coworkers did much work and had a deep insight into DDT process [2-4]. Excellent reviews that summarize our understanding to date have been given by Lee and Moen [5] Shepherd and Lee [6], and Zhao et al [7].

In the experiments, the flame is usually accelerated by putting obstacles or spiral-coils into smooth tubes, and turbulence plays an important role in DDT [2-4, 8-11]. But, the mechanism of the inherent threshold in DDT is not known to date. It should be noted that the DDT process is very complex and transient, and the complex mechanism cannot be observed in experiments. Numerical simulations have been done extensively in recently years [12-16], but the mechanism is still unknown. Since the flame acceleration involves all mechanisms that are sensitive to different initial and boundary conditions, it seems nearly impossible to have a universal theory to describe the DDT process [15].

It is a common point of view that the obstacles create turbulence and the turbulence enhances the coupling between turbulent flame and incident shock wave. Lee proposed a mechanism of Shock Wave Amplification by Coherent Energy Release (SWACER) [17] to explain this phenomenon. It can be observed both in experiments and numerical simulations that the DDT process always occur when the velocity of deflagration reaches about 50% C-J detonation velocity, which is about the sound speed of the combustion products [18-23]. No deflagration wave with a speed faster than 50% C-J detonation velocity has been observed in practice. It is obviously that this is a critical state.

The quantitative prediction of DDT is very important to industries because detonation has very destructive power. It is one of the major unsolved problems in combustion and detonation theory. It is also an extremely interesting and difficult scientific problem because of the complex nonlinear interactions among the different contributing physical processes, such as turbulence, shock interactions, and energy release. It is not surprising, therefore, that a quantitative criterion of DDT process has not been yet derived. Such a criteria is necessary for various estimates of the explosion hazard of combustible systems in practice. In this paper, the DDT process is theoretically analyzed and the theoretical expression about this critical condition is given.

## 2. Theoretical Analysis

The DDT process in a smooth tube is analyzed in this paper. The structure of the deflagration wave is a leading shock wave (SW) followed by a flame surface. Figure 1 gives the structure of a deflagration wave in the laboratory coordinate. The flow field is divided into three regions by the SW and flame surface. Region 1 is the premixed detonable mixture at initial pressure and temperature at rest. Region 2 is the preheated mixture behind the SW, and region 3 is the combustion products.

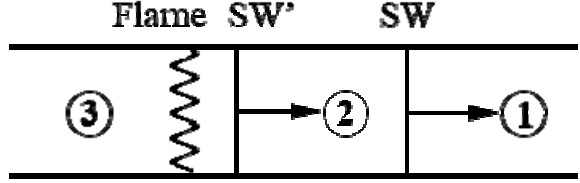

**Fig. 1.** Structure of a deflagration wave in the laboratory coordinates

This is a weak DDT process. The flame is accelerated from laminar flame to turbulence flame gradually and slowly. Suppose the temperature of the preheated mixture in region 2 is much lower than the auto-ignition temperate and the combustion only takes place on the flame front. No auto-ignition or hot spot occurs in region 2.

At the beginning stage, the leading shock wave is very weak and the flow behind it is subsonic. Therefore, the small pressure rise caused by weak combustion can propagate upstream by sonic waves and enhance the strength of the SW. With the strength of leading shock wave SW becoming stronger, the combustion behind it becomes stronger. The big pressure rise caused by violent combustion can produce a series of secondary shock waves (SW'), which is shown in Fig.1. The strength of SW' is a function of pressure rise on flame front and the parameters in region 2.

It should be noted that the strength of the leading shock wave SW cannot be enlarged without limit. It is limited by the energy release of combustion on the flame front. If the Mach number of SW is too high, the secondary shock wave SW' will become weaker. There is a negative feedback mechanism between the leading shock wave and the flame surface.

Suppose there exist a critical state that when the secondary shock wave SW' catches up and merges with the leading shock wave, the thermodynamic parameters of the new shock wave are exactly equal to the thermodynamics parameters of a C-J detonation, and therefore, a DDT process occurs. We will analyze the thermodynamic characteristics of this critical state by using theories of shock wave dynamics in the following part.

In order to simplify the analysis process, the specific heat ratio of premixed detonable mixture is assumed to be $\gamma=1.4$ and keeps constant. According to the normal shock wave relations, the parameters in region 2 can be calculated by Eqs.(1) and (2),

$$\frac{p_2}{p_1} = \frac{2\gamma M_1^2 - (\gamma-1)}{\gamma+1} = \frac{7M_1^2 - 1}{6} \qquad (1)$$

$$\frac{T_2}{T_1} = \frac{\left[2\gamma M_1^2 - (\gamma-1)\right]\left[(\gamma-1)M_1^2 + 2\right]}{(\gamma+1)^2 M_1^2} = \frac{(7M_1^2 - 1)(M_1^2 + 5)}{36M_1^2} \quad (2)$$

where, $M_1$ is the Mach number of the leading shock wave, $\gamma$ is the specific heat ratio, $p_1$ and $T_1$ are the initial pressure and temperature in region 1, respectively. And $p_2$ and $T_2$ are the pressure and temperature in region 2, respectively.

In order to determine the pressure rise after combustion, we suppose that the time scale of heat release is very shot and can be neglected compared with the flow time scale, and the combustion process is a constant-volume combustion process. Therefore, the pressure $p_3$ can be calculated by Eq.(3),

$$\frac{p_3}{p_2} = \frac{T_0}{T_2} \quad (3)$$

where, $T_0$ is the total temperature of constant-volume combustion.

In addition, we choose the pressure as the controlling parameter of the critical state, the relationship is given by Eq.(4),

$$\frac{p_3}{p_1} = p_{\text{ZND}} \quad (4)$$

where, $p_{\text{ZND}}$ is the nondimenstional von Newman spike of a C-J detonation.

Combining the Eqs.(1)-(4), we can easily obtain the theoretical results for this critical condition by Eq.(5),

$$\frac{6M_1^2}{M_1^2 + 5} = p_{\text{ZND}} \frac{T_1}{T_0} \quad (5)$$

From Eq.(5), we can find that, for a C-J detonation whose von Newmann pressure spike is certain, the Mach number of the deflagration wave is only determined by the combustion energy release. This is the key mechanism of the DDT process. The von Newmann spike can be calculated by Eq.(6),

$$p_{\text{ZND}} = \frac{7M_{\text{CJ}}^2 - 1}{6} \approx \frac{7M_{\text{CJ}}^2}{6} \quad (6)$$

Finally, we obtain the critical criterion for DDT process by Eq.(7),

$$\frac{6M_1^2}{M_1^2 + 5} = \frac{7M_{\text{CJ}}^2}{6} \frac{T_1}{T_0} \quad (7)$$

In the following part, we will give two examples to demonstrate this theoretical result briefly. The first example is from reference [19], where the detonable mixture is a $CH_4+2O_2$ mixture at 8.2kPa and 300K. The comparison results are given in Table 1 and Fig.1. The second example is from reference [23], where the detonable mixture is stoichiometric ethylene-air mixture at 100kPa. The comparison results are given in Table 2 and Fig.2. We can find that the theoretical results are in excellent agreement with experimental results. At the critical condition, the velocity of the leading shock wave is about 60% of C-J detonation velocity, and it is also close to the sound speed of the combustion products.

**Table 1.** Parameters of DDT process of $CH_4+2O_2$ [19]

| Parameters | Values |
|---|---|
| $M_1$ | 4.33 |
| SW(m/s) | 1546.6 |
| $P_1$(kPa) | 8.2 |
| $T_1$(K) | 300 |
| $a_1$(m/s) | 356.7 |
| $P_2$(kPa) | 184.5 |
| $T_2$(K) | 1067.1 |
| $P_{ZND}$(kPa) | 453.4 |
| $T_0$(K) | 3280 |
| $M_{CJ}$ | 6.71 |
| $D_{CJ}$(m/s) | 2395.1 |
| $a_{CJ}$(m/s) | 1337.6 |
| SW/$D_{CJ}$ | 64.5% |

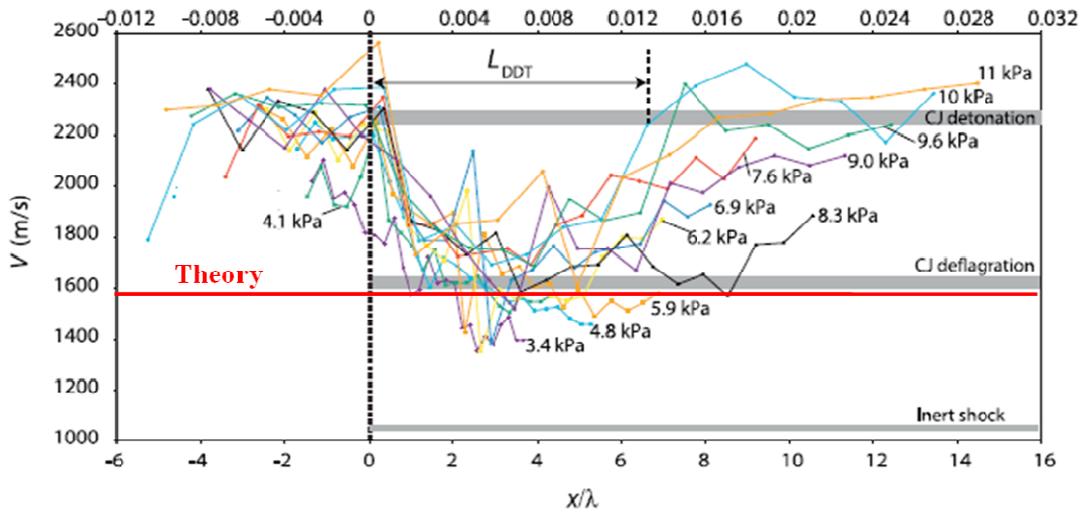

**Fig.1.** Comparisons of theoretical results with experimental results of $CH_4+2O_2$ mixture [19]

Table 2. Parameters of DDT process of stoichiometric ethylene-air mixture [23]

| Parameters | Values |
|---|---|
| $M_1$ | 3.16 |
| SW(m/s) | 1093.2 |
| $P_1$(atm) | 1.0 |
| $T_1$(K) | 300 |
| $a_1$(m/s) | 346.0 |
| $P_2$(atm) | 11.65 |
| $T_2$(K) | 800.3 |
| $P_{ZND}$(atm) | 33.53 |
| $T_0$(K) | 2700 |
| $M_{CJ}$ | 5.27 |
| $D_{CJ}$(m/s) | 1825.9 |
| $a_{CJ}$(m/s) | 1041.7 |
| SW/$D_{CJ}$ | 59.9% |

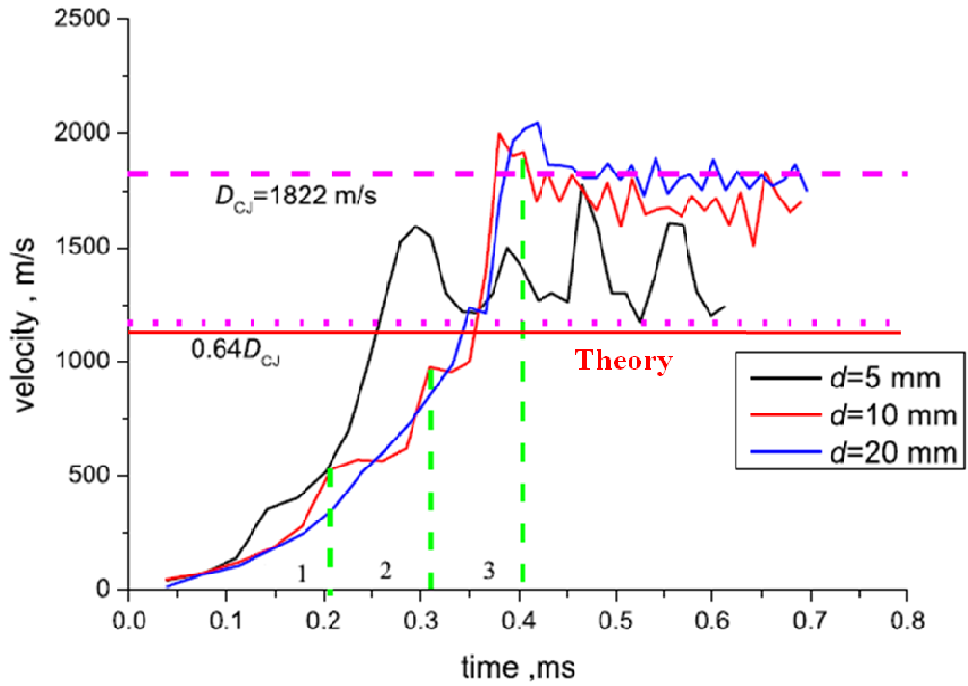

Fig.2. Comparisons of theoretical results with experimental results of ethylene-air mixture [23]

## 3. Conclusion

In this study, the physical model of deflagration-to-detonation transition is put forth, the mechanism is analyzed and the theoretical criterion is derived. The theoretical criterion is $\dfrac{6M_1^2}{M_1^2+5} = \dfrac{7M_{CJ}^2}{6}\dfrac{T_1}{T_0}$. When the Mach number of the deflagration wave reaches this critical value, a C-J detonation will be triggered immediately and DDT process occurs. This is also the maximum propagating velocity of a deflagration wave. This analysis reveals that a C-J detonation can be considered as a superposition of two shock waves propagating in the same direction and this is a unique solution.